\documentclass[12pt]{article}
\usepackage{amsmath}
\usepackage{amssymb}
\usepackage{epsfig}
\usepackage{cite}
\usepackage{color,colordvi}
\newcommand{\eq}{\begin{equation}}
\newcommand{\eqx}{\end{equation}}
\newcommand{\eqn}{\begin{eqnarray}}
\newcommand{\bi}{\begin{itemize}}
\newcommand{\eqnx}{\end{eqnarray}}
\newcommand{\ei}{\end{itemize}}

\newcounter{hran}

\def\MSbar{\relax\ifmmode\overline{\rm MS}\else{$\overline{\rm MS}${ }}\fi}

\def\simgt{\stackrel{>}{{}_\sim}}

\begin{document}
\begin{titlepage}
\begin{center}

\hfill CERN-PH-TH/2010-226
\vskip 0.4 cm

{\Large\bf UV-Completion by Classicalization}

\vspace{0.2cm}

\end{center}

\begin{center}

{\bf Gia Dvali}$^{a,b,c,d}$, {\bf Gian F. Giudice}$^{c}$,  {\bf Cesar Gomez}$^{e}$ and 
{\bf Alex Kehagias}$^{f}$


\vspace{.2truecm}

{\em $^a$Arnold Sommerfeld Center for Theoretical Physics,
Fakult\"at f\"ur Physik\\ Ludwig-Maximilians-Universit\"at M\"unchen,
Theresienstr.~37, 80333 M\"unchen, Germany}


{\em $^b$Max-Planck-Institut f\"ur Physik,
F\"ohringer Ring 6, 80805 M\"unchen, Germany}


{\em $^c$CERN,
Theory Division,
1211 Geneva 23, Switzerland}


{\em $^d$CCPP,
Department of Physics, New York University\\
4 Washington Place, New York, NY 10003, USA}


{\em $^e$Instituto de F\'{\i}sica Te\'orica UAM-CSIC, C-XVI \\
Universidad Aut\'onoma de Madrid,
Cantoblanco, 28049 Madrid, Spain}\\

{\em $^f$Physics Division, National Technical University of Athens, \\
15780 Zografou Campus, Athens, Greece} \\

\end{center}

\vskip .2in

\setlength{\baselineskip}{0.2in}

\centerline{\bf Abstract}
\vskip .1in

\noindent


 We suggest a novel approach to UV-completion of a class of non-renormalizable theories, 
 according to which the high-energy scattering amplitudes get unitarized by production of extended classical objects 
({\it classicalons}), playing a role analogous to black holes, in the case of non-gravitational theories.  
The key property of classicalization is the existence of a   
classicalizer field  that couples to energy-momentum sources. Such localized sources are excited in high-energy scattering 
processes and lead to the formation of classicalons. Two kinds of natural  classicalizers are 
Nambu-Goldstone bosons (or, equivalently, longitudinal polarizations of massive gauge fields) and scalars coupled to 
energy-momentum type sources.  Classicalization has interesting phenomenological 
applications for the UV-completion of the Standard Model both with or without the Higgs.  
In the Higgless 
Standard Model the  high-energy  scattering amplitudes of longitudinal $W$-bosons self-unitarize via classicalization,
 without the help of any new weakly-coupled physics. Alternatively, in the presence of a Higgs boson, classicalization could explain the stabilization of  the hierarchy.   In both scenarios the high-energy scatterings 
are dominated by the formation of classicalons, which subsequently decay into many particle states.  The experimental signatures at the LHC are quite distinctive, with sharp differences in the two cases.

\end{titlepage}

\vskip .4in
\noindent


\section{Introduction}

   In this note, we wish to suggest an alternative approach to  UV-completion of some non-renormalizable theories, which we shall refer to as {\it classicalization}.  
  In order to outline the essence of this phenomenon, consider a non-renormalizable theory  
in which particle interactions are governed  by a coupling constant  of dimensionality of an inverse mass, and let us study a two-to-two scattering process in such a theory.  
The perturbative amplitude grows as a positive power of the kinematic parameter $s$ and violates unitarity above a certain 
scale $M_*\, \equiv \, L_*^{-1}$.  The standard (Wilsonian) approach to the problem is to build a UV-completion 
by integrating-in some new degrees of freedom that reconstruct a weakly coupled  quantum field theory above the scale $M_*$.   

We shall suggest a different path, according to which the violation of unitarity is only an artifact of the perturbative approach, and instead the theory self-unitarizes  by classicalization.   In order to understand how this could happen, 
imagine that scattering of particles  at distance $1/\sqrt{s}$ requires exciting a source $J(s)$ that grows as a positive power of $s$. Such source produces a strong classical field 
$\phi$ within a region of effective size $r_*(s)$, which grows as some positive power of $s$. 
This means that the scattering is accompanied by formation of a classical configuration of 
$\phi$,   a {\it classicalon}.   We shall refer to the scale $r_*(s)$, setting the size of the classicalon, as 
the classicalization radius, or $r_*$-radius for short.  As a result  of this phenomenon, the  scattering process  classicalizes.    

In the perturbative approach, the two-to-two scattering
is accompanied by hard processes with momentum  transfer of the order of $\sqrt{s}$, which apparently violate unitarity.   However, in reality,  since the decay of a classical object into  any two-particle state is exponentially suppressed,   the two-to-two scattering amplitude with high-momentum transfer must be exponentially suppressed,  
\begin{equation}
A_{2\rightarrow 2} \, \sim \, {\rm e}^{-(\sqrt{s}/M_*)^c} \, , 
\label{theA}
\end{equation}
 where $c$ is some positive model-dependent number.  Instead,  at high $s$, the leading  contribution to two-to-two scattering processes will come from soft scattering with momentum transfer $\sim r_*(s)^{-1} \, \ll \, M_*$,  which is automatically below the unitarity bound.  
 
At high $s$, the scattering becomes dominated by production of long-lived classicalons  of size $r_*(s)$, with geometric cross-section 
\begin{equation}
 \sigma \, \sim  \,  r_*(s)^2 \, .
 \label{geo}
 \end{equation}
As mandatory for classical configurations, the classicalons slowly  decay into many light quanta over time-scales $t_{class} \,  >   r_* \gg 1/M_*$.   The physics of such objects 
 is entirely dominated by properties of the theory at long distances.  
  {\it In the other words,  classicalization converts the high-energy physics into a long distance physics.} 
  
 The key feature of classicalizing theories is the $r_*$-phenomenon,  according to which the  theory responds to localized sources by creating large classical objects.  
 A famous example of   $r_*$-phenomenon is the Schwarzschild radius of black holes in gravity. 
Another important example of $r_*$-phenomenon, very relevant for our discussion, is the creation
of  $r_*$-radius classicalons by  longitudinal (St\"uckelberg) components of the interacting massive vector fields, 
in the absence of the weakly-coupled Higgs degree of freedom \cite{dhk}.

   Notice that, classicalons need not be  classically-stable configurations. Nor they must correspond to any topological solitons.   Although, as we shall explain,  sharply localized topological charges  can also serve as  sources for large-size classicalons,  
the production rate of topological charges  in two-particle scatterings is exponentially suppressed  due to the required coherence.   As a result, they cannot  play any efficient role in unitarizing the scattering amplitudes.   

   Classicalons  that provide UV-completion of the theory  must be sourced  by  Noether
(as opposed to topological) charges, such as energy and  momentum.   Such charges leave Gaussian fluxes at infinity that  guarantee the formation of classicalons. 
This is the key to why production of such classicalons is the dominant effect in high-energy scattering.

   An  example of a universal classicalizer is gravity.  Consider for instance a process 
   of two-to-two graviton scattering at  distance $1/\sqrt{s}$ and energy $\sqrt{s}$.  The tree level amplitude of this process grows as $\sim \alpha_{gr}(s)\, = \, G_Ns$, where $G_N \equiv M_P^{-2}$ is the Newton's constant.   Violation of unitarity in this process is due to the fact that the 
effective   gravitational coupling grows at short distances and blows up for $s \, \gg \, M_P^2$. However, this violation is an artifact of perturbation theory.  We know that the localization of a particle 
pair of center-of-mass energy $\sqrt{s} \,  \gg \, M_P$ is impossible at distances shorter than the Schwarzschild  radius of corresponding  mass, $R_s \equiv \, 2G_N\sqrt{s}$.  This fact is insensitive to  the short distance properties of the theory, and suggests the way out of  the seeming violation of unitarity in  two-to-two scattering.    
 As is well understood, the two-to-two scattering for $\sqrt{s} \, \gg \, M_P$ is exponentially 
 suppressed.  Instead, the scattering process is dominated by production of classical black holes 
 that decay slowly by Hawking evaporation into many quanta.  In other words the theory classicalizes 
 itself by black hole formation. 
 
  Based on the above universal classicalizing property of gravity,  in ref.~\cite{giacesar} it was suggested
 that the scattering of longitudinal $W$-bosons in the Higgless  Standard Model can also be unitarized via  classicalization.  The idea is that in the presence of extra spin-2 and spin-0 degrees of freedom that  get strong couplings to energy-momentum sources at the scale   
$M_*\sim$~TeV,   $WW$-scattering must get classicalized via black hole formation for $\sqrt{s} \gg $TeV.  One of the questions that we ask in the present work is: how essential is the 
existence  of extra degrees of freedom for achieving the classicalization effect in the Standard Model?  
 In other words, the question is whether the Standard Model itself possesses the resources for unitarizing 
$WW$-scattering via classicalization.   
   
 Our goal is to introduce the concept of classicalization in non-gravitational theories.  
 Although the analogy with black holes serves as a source of inspiration for our idea, gravity is not the main focus of our paper. On the contrary, we depart from gravitational systems as much as possible, because we want to disentangle the classicalization phenomenon from the peculiarities of gravity.  The goal is to reduce the analysis to its bare essentials, when energy (self)sourcing results into the appearance of a macroscopic classical length-scale governing the interaction range at high energies, and to show that these  bare essentials  are shared by a class
of non-gravitational systems.  From this point of view, otherwise important concepts such as entropy and thermal properties are only peculiarities of the way
gravity accommodates the existence of the horizon, but are not essential for unitarity. Any consistent theory in which the high-energy scattering of two particles is dominated by a multi-particle final state obtained through an intermediate classical configuration is expected to unitarize.
 The essence of the phenomenon is that a seeming violation of unitarity is cured  by production of classical objects at very high energies.   In a certain sense, classicalization is a ``self-defense"
 of the theory against the absence of a Wilsonian UV-completion. 
  The purpose of this note is not to provide a complete proof of classicalization, but to suggest 
  the key idea and provide some crucial evidence. 
 
  As we shall show,  there are two phenomenologcally-important categories of low spin particles that exhibit classicalization: 
 
      {\it  1) Nambu-Goldstone bosons  (or, equivalently, longitudinal components of massive vector gauge  fields);

    2) Higgs-like scalars.} 
   
  The first category  of fields is automatically  sourced by the derivative energy-momentum interactions. In the absence of a weakly-coupled Higgs degree of freedom, these fields are known to exhibit the
$r_*$-phenomenon,  responding by creating an extended classical configuration around a localized source \cite{dhk}.  
 Thus, as we shall see, these theories have the tendency to 
  self-classicalize.   An obvious phenomenological application of this phenomenon is the
  unitarization of the high-energy $WW$-scattering amplitudes in the {\it Higgless}  Standard Model by means of  classicalization of longitudinal $W$-bosons.  Needless to say, this would be an astonishing possibility, 
  suggesting that that Standard Model self-completes itself by classicalization without any need of
  a Higgs particle (or any other weakly-coupled physics\cite{weakhiggs}).    
   
     In the second case, classicalization is not automatic, but takes place whenever 
  the scalar in question is sourced by the energy-momentum.    This offers a possibility to address the hierarchy problem in the Standard Model (with the Higgs), by  using the Higgs field as a classicalizer.   
      
  Below we shall explore both possibilities.  A characteristic feature in both cases is that the  relevant 
 two-to-two scattering amplitudes at high $\sqrt{s}$ saturate at relatively  low momentum  transfer, while the scattering is dominated by production of classicalons that  decay 
 into many-particle final states. 
 
  An important difference is that, in the Higgless Standard Model, this only applies to processes  
  that involve  $WW$-scattering and the Higgs particle is replaced by classicalon resonances.  On the other hand, when the Higgs behaves as classicalizer, the feature of classicalon-production is universal for all high-energy collisions, but a  weakly-coupled Higgs must exist with mass below the classicalization scale.  
  
  Finally, classicalization of  non-gravitational theories can represent a simplified  laboratory 
  in which one can test ideas about  self-completeness of gravity~\cite{giacesar}. The unitarization of high-energy amplitudes is reduced to its bare essential of creating a classical object, and one can decouple from the problem all the
subtleties   connected with the existence of horizons or the complications related to information loss.

\section{The Essence of Classicalization}

   The phenomenon of {\it classicalization}  is  a non-Wilsonian self-completion of the theory by localized particles becoming  classical in deep-UV \cite{giacesar}.   As we have argued earlier, in consistent gravity theories clasicallization appears to be a built-in property, because of black holes, which prevent  us from performing measurements at  arbitrarily short distances.   But the notion of classicalization can be generalized to a class of non-gravitational non-renormalizable  theories. 
    
      Of course, in non-gravitational theories there are no black holes, so strictly speaking it is always possible to retrieve  information from arbitrarily short scales if one uses arbitrary external sources.  However, if we switch-off gravity, then a complete inaccessibility  of information is  unnecessary 
 for the UV-completion of a particular interaction.   Instead, it is sufficient that 
 the storage of a {\it particular type}  of  information in sources  relevant for particularly badly behaved correlators or amplitudes  results in the formation of extended classical objects.   
  
   For example,  consider a theory in which a bosonic field  $\phi$ is sourced by an operator $J$, through an interaction
   $\phi J$, where $J$ is some operator  that may depend on $\phi$ and also on some other fields. 
   Imagine that, at the perturbative level, the theory becomes sick at high energies because of the  bad behavior of correlators such as $\langle JJ\rangle$. 
 If these correlators  involve integration 
   over sources that produce {\it classical} configurations of $\phi$, the bad behavior becomes  an artifact of the perturbative expansion. Instead, if we take into the account that those integrands  are in reality classical objects, integration over such sources would be exponentially suppressed. 
   This is the key point of classicalization. 
   
    For the classicalization phenomenon the essential ingredient is a {\it classicalizer} field ($\phi$) that  generalizes the role of gravity  in creating the extended objects.  The key point is that 
   the field $\phi$ must be sourced by an operator ($J$) that becomes strong whenever   
particles are localized at short distances. In this way, localization of particles is accompanied by 
formation of a strong classical field $\phi$, so that short distance scattering is accompanied 
by creation of extended classical objects.   The characteristic distance scale that serves as a measure of classicality of a  given source will be called here the associated $r_*$-radius, and it 
is a generalization of the Schwarzschild radius for non-gravitational theories. 
Essentially the $r_*$-radius measures the effective extent of the strong classical field $\phi$ around the source.

 Schematically the phenomenon of classicalization can be understood in the following general terms. Let us think in terms of a generic Lagrangian, 
 \begin{equation}
 {\cal L}(\phi,J) = {\cal L}(\phi) + {\phi \over M_*} J \, ,
 \label{clas1}
 \end{equation}
  where $J$ is a source. This Lagrangian is such that the two-to-two perturbative scattering amplitude $A(s,t)$ for $\phi$ quanta violates unitarity at the UV length-scale $L_{*}$, {\it i.e.} $A(1/L_{*}^{2}) =O(1)$. 
  For example, an  effective Lagrangian (for the simplest case of spin-zero $\phi$ quanta)  leading to this violation of unitarity can be of the form, 
 \begin{equation}
 {\cal L}(\phi)= (\partial_{\mu}\phi)^2  +\frac{1}{M_{*}} \phi(\partial_{\mu}\phi)^2
 +\frac{1}{M_{*}^5} \phi(\partial_{\mu}\phi)^4 \, + \, ...
 \label{cubic}
 \end{equation}
 with $L_{*}=M_{*}^{-1}$ . 
 The non-linear terms in the above Lagrangian provide an example of  self-sourcing of the 
$\phi$ field. The source at the perturbative level leads to violation of unitarity  for 
energies $\sqrt{s} \, \gg \, M_* $, but we wish to argue that the same interaction at the non-perturbative level  restores unitarity by classicalization. 
 
Classicalization takes place whenever the strength of  a source localized within a region of size 
$L$ grows as a positive  power of $1/L$.  In particular, this  is the case for the energy-momentum  type sources, and this property is shared by the cubic self-coupling in eq.~(\ref{cubic}).  For this reason, the 
scattering of $\phi$ particles must classicalize at high energies.  In order to see this, let us consider the scattering of wave packets of the $\phi$ field at some center-of-mass energy $\sqrt{s} \, \gg \, M_*$. In order to scatter with momentum transfer $1/L$, the particles have to come within a distance $L$.  Perturbatively it seems that, as long as $L \, > \, L_*$, such localized wave-packets are well-described quantum states.  However,  this is not the case.  In fact,  for  the wave-packets  
that are localized within the distance    
 \begin{equation}
 \label{rrs}
 r_*(s) \sim L_{*}^2 \sqrt{s} \, ,
 \end{equation}
physics enters in the classical regime by forming a classical configuration of radius $r_*$.  In other words, the system classicalizes.   Thus, due to self-sourcing, the  localization of quanta 
of center-of-mass energy $\sqrt{s} \, \gg \, M_*$ is impossible within a distance shorter 
than the corresponding $r_*$-radius given by eq.~(\ref{rrs}). 
In order to prove that this is the case, let us assume the opposite. Imagine that we localized 
particles of center-of-mass energy $\sqrt{s} = 1/L \, \gg M_*$  within a sphere 
of radius $R \, \ll r_*$. Because of the cubic self-interactions, such localization produces 
an effective localized source for the $\phi$-field with integrated value $J \sim 1/LM_*$. 
Thus,  far from the localization region, the dynamics of the weak field $\phi$ 
should be well-described by the following  linearized equation,\footnote{Here and in the following $\delta(r)$ represents the three-dimensional Dirac delta-function $\delta^{(3)}({\vec r})$. For our purposes, it is convenient to use the representation 
$$ \delta(r)=\frac{1}{4\pi}\frac{d}{dx_i}\left( \frac{x_i}{r^3}\right),$$
where $x_i$ are the three-dimensional space coordinates and $r=(x_i^2)^{1/2}$.} 
 \begin{equation}
 \Box \,  \phi \, +...\, = \, \delta(r)  {L_*\over L} \, .
 \label{gauss1}
 \end{equation}
 This equation shows  that, since $\phi$ is sourced by the energy,  the localization 
 of quanta implies the existence of a Gaussian flux of the gradient ${\vec \nabla}\phi$ at infinity,  and  $\phi$ has to  grow as 
 \begin{equation}
 \phi(r) \, \sim \, \frac{L_*}{r L}\, ,
 \label{phi}
 \end{equation}
 as we approach the localization region.  Notice, that non-linear interactions cannot prevent this  growth until $\phi$ becomes of order $M_*$, which  happens at the distance-scale 
  \begin{equation}
 \label{two}
 r_*(L) \sim \frac{L_{*}^2}{ L} \, .
 \end{equation}
Thus,  we see that  a source of energy  $\sqrt{s} = 1/L \gg M_*$  acquires an effective size 
$r_* \, \gg \, L_*$ and becomes a classical object.  

 The result is that, although in the naive perturbative approach one would think that high-energy scattering at $\sqrt{s} \gg M_*$ is dominated by hard-collisions  with momentum 
 transfer $\sim \sqrt{s}$, in reality it is totally dominated by formation of classical objects 
 of radius  $r_* =  L_*^2 \sqrt{s}$ that eventually decay into many particles.
  Two-to-two particle scattering amplitude  at high momentum transfer is  exponentially suppressed,  and instead is dominated by soft processes with momentum transfer as low as $\sim 1/r_*$.


It is very important to note that, although classicalization of sources at $\sqrt{s} = 1/L \, \gg \, M_*$ 
is an ultra-high-energy phenomenon, it 
is entirely dominated by long-distance dynamics corresponding to momenta 
$1/r_* \, \ll \, M_*$.  Because of this, all the possible higher-order corrections are under control.  
 In other words, the  higher-order terms in the Lagrangian,  
\begin{equation}
{1\over M_*^{n+4k-4}}\phi^n (\partial_{\mu}\phi \partial^{\mu}\phi)^k \, ,
\label{higherterms}
\end{equation}
 are not changing the dynamics of classicalization.   This can be seen from the following 
 analysis.   The asymptotic 
 behavior at large $r$ is  $\phi(r) \sim  (M_*Lr)^{-1}$, meaning that at large distances the terms (\ref{higherterms})   behave as 
 \begin{equation}
{1\over M_*^{n+4k-4}}\phi^n (\partial_{\mu}\phi \partial^{\mu}\phi)^k \,  \sim \, 
(M_*^{2n+6k-4}L^{2k+n}r^{4k+n})^{-1} \, .
\label{higherterms1}
\end{equation}
The radius  $r_{n,k}$ for which a given  non-linear term  becomes of the order of the linear term ($n=0, k=1$)  is
\begin{equation}
 r_{n,k} \, = \, L_* \left(\frac{L_*}{L}\right)^{{2k+n-2 \over 4k+n-4}} \, .
 \label{rnk}
 \end{equation}
The  $r_*$-radius in eq.~(\ref{two}) corresponds to the largest of these scales. 
Indeed, for $k=1$, any $r_{n,1}$ is parametrically equal to $r_*$. The addition of extra derivatives ({\it i.e.} increasing the value of $k$) reduces $r_{n,k}$, while adding powers of fields ({\it i.e.} increasing $n$) increases $r_{n,k}$. However, for $k>1$, any $r_{n,k}$ is smaller than $r_*$. Thus, ignoring accidental cancellations, we conclude that the largest $r_*$-radius is set by eq.~(\ref{two}). In the case of a purely polynomial interaction ($k=0$), the leading effect comes from the operator with the smallest number of fields and thus the exponent in eq.~(\ref{rnk}) ranges between 1 and 3.
 In general,  the scale $r_*$ is determined by the leading non-linearity, whereas 
higher non-linearities only affect the behavior of  $\phi$ at short distances $r \, \ll \, r_*$. Because of this,  they cannot affect the size of the classical configuration and they do not play an
 important role in the scattering processes at large $s$. 
 
 
   \subsection{An Example}
  
   In order to demonstrate explicitly the insensitivity of the classicalization phenomenon 
   with respect to the higher-derivative interactions, consider a scalar field with DBI-type action,
 \begin{equation}
 M_*^4\sqrt{1 \,  + \, L_*^4 (\partial_{\mu}\phi)^2  }  \, + \, {\phi \over M_*}  J\, .
 \label{dbiscalar}
 \end{equation}
 We assume that the scalar $\phi$ transforms under a shift symmetry as $\phi \,  \rightarrow \, \phi \, + \, c$,  which (up to boundary terms) forbids any non-derivative interaction of $\phi$.  
Let us consider the response to a localized spherically-symmmetric source 
$J(r)  = \delta(r) 4\pi / L$, with 
$1/L \, \gg \, M_*$.  For such a source, 
the equation of motion, 
\begin{equation}
\partial^{\mu} \left ( {\partial_{\mu} \phi  \over \sqrt{1 \,  + \, L_*^4 (\partial_{\mu}\phi)^2}  } \right )
 \,  = \,  {J \over M_*} \, ,
\label{equation}
\end{equation}
can be easily recast as an algebraic equation for  $\partial_r\phi(r)$, 
\begin{equation}
 {\partial_{r} \phi  \over \sqrt{1 \,  -\, L_*^4 (\partial_{r}\phi)^2}  } 
 \,  = \, -\frac{L_*}{r^2 L} \, .
\label{equation1}
\end{equation}
Its solution is 
 \begin{equation}
 \partial_r\phi(r) \, = \,  M_*^2{1 \over  \sqrt{1 \, + \, (r/r_*)^4}} \, ,
 \label{rootdbi2}
 \end{equation}
  where $r_* = L_* \sqrt{L_*/L}$.  The action (\ref{dbiscalar}) includes  only the invariants 
 that contain exactly one derivative per $\phi$.

  Let us now supplement this action by 
  arbitrary operators that contain more derivatives per $\phi$,
 \begin{equation}
 M_*^4\sqrt{1 \,  + \, L_*^4 (\partial_{\mu}\phi)^2  }  \, + 
\, M_*^4 \, \sum_{n>2,k>0} \, I_{n,n+k}\left[ \partial^k,  (\partial \phi)^n\right] \, +\,   {\phi \over M_*}  J\, ,
 \label{dbiscalar1}
 \end{equation}
 where $I_{n,n+k}[\partial^k,  (\partial \phi)^n]$ schematically denotes an invariant that contains
 $n$ powers of $\phi$ and $n+k$ powers of the derivative.  
  Since according to eq.~(\ref{rootdbi2}) the behavior of  $\partial_r\phi$ is 
  \begin{equation}
 \left. \partial_r\phi(r) \right|_{r \gg r_*} \,  \sim \,  M_*^2 \, \frac{r_*^2}{r^2} ~~~{\rm and} ~~~
  \left.\partial_r\phi(r) \right|_{r \ll r_*} \,  \sim \,  M_*^2 \, \left( 1 \, - \, \frac{r^4}{2  r_*^4} \right)\, ,
  \label{large and small}
  \end{equation}
  the derivatives behave as
 \begin{equation}
  \frac{\partial^k}{M_*^k} \left. \left[ \partial \phi(r) \right]^n \right|_{r \gg r_*} \,  \sim \, \left[ \partial \phi(r) \right]^n   {L_*^k\over r^k} ~~~{\rm and} 
~~~
  \frac{\partial^k}{M_*^k} \left. \left[ \partial \phi(r) \right]^n \right|_{r \ll r_*} \,  \sim \, \left[ \partial \phi(r) \right]^n   {{L_*^k}\over {r_*^4\, r^{k-4}}}\, .
  \label{large and small2}
  \end{equation}
Thus, higher derivatives are negligible as long as $r \, \gg \,  L_*$.     
  
  In other words,  
  adding extra $n$-derivatives to any particular invariant amounts to lowering the value 
  of that invariant at least by $(L_*/r)$. Thus, any such invariant will be subdominant with respect to 
  the action (\ref{dbiscalar}), for $r \gg L_*$. In particular, higher-derivative terms will play no role in the formation of the classicalon at $\sqrt{s} \, \gg \, M_*$.   
 
  The insensitivity of classicalization  to higher-order invariants is completely analogous to the case of
 gravity, in which  the higher-order curvature  invariants play no role in the formation and evolution of classical black holes of size larger than the Planck length, $L_P$. In our case, 
  the role of $L_P$  is played by the scale $L_*$, whereas $r_*$ takes the role of the Schwarzschild radius.  
    The analogy with gravitational self-completeness is already clear from eq.~(\ref{two}). In fact eq.~(\ref{two}), for $L_{*} = L_P$, is at the core of the bouncing of transplanckian configurations into classical black holes.

 Classicalization of non-unitary theories is therefore crucially dependent on the nature of the source. As it is clear from eq.~(\ref{gauss1}), in order to classicalize processes at high momentum transfer $t$,   the source of $\phi$ quanta should have the property to increase with $t$, and   
should   behave as an effective  energy source with energy $1/L$ set by the localization width. Interestingly enough there are non-gravitational theories where such a behavior appears quite naturally. 
A well-known  case in which the spin-zero $\phi$ quanta necessarily have derivative couplings to the source occurs for Nambu-Goldstone bosons  and consequently  for longitudinal (St\"uckelberg) components of the gauge fields.  
We shall provide evidence that interactions of such particles have the tendency to classicalize. 
This opens the possibility of 
self-completeness, by means of classicalization, of theories with massive gauge fields and/or  non-linear sigma-models. 

\subsection{Quantum and Classical Scales}

Before turning to a more detailed analysis of the phenomenon of classicalization, we would like to comment on the physical meaning of the various length scales. 
Let us consider the Lagrangian
  \begin{equation}
 {\cal{L}}=\frac 12 (\partial_\mu\phi)^2+G_\phi\,  {\cal O}_{n,k}\, .
\end{equation} 
Here $ G_\phi = M_*^{4-n-2k}$ and ${\cal O}_{n,k}\sim \partial^{2k}\phi^n$ is a generic non-renormalizable operator containing $n$ powers of the field and $2k$ derivatives (with $n> 2$ and $n+2k>4$). Of course Lorentz invariance always requires an even number of derivatives.  

We are considering the case of self-sourcing, in which the classicalizer acts as a source of its own field. We are interested in the scattering process of $\phi$ quanta at large center-of-mass energy and large momentum transfer. But, for simplicity, we will treat only the static case in which we localize within a volume of radius $r_*$ an amount of energy $\sqrt{s}$ in the $\phi$ field. Parametrically, this corresponds to
\begin{equation}
\sqrt{s} \sim \int_{r<r_*}d^3r \, (\partial \phi )^2 \sim r_* \, \phi^2 \, .
\end{equation}
This localized wavepacket acts as a source for the field $\phi$ itself with an integrated value parametrically given by
\begin{equation}
Q=\int_{r<r_*}d^3r \, G_\phi \, \frac {\delta {\cal O}_{n,k}}{\delta \phi} \sim G_\phi \, r_*^{3-2k}\, \phi^{n-1} \sim \left( M_* r_*\right)^{\frac{7-n-4k}{2}}\left(\frac{\sqrt{s}}{M_*}\right)^{\frac{n-1}{2}}.
\end{equation}

From the previous examples we have learned that, far from the localization region, the field behaves as $\phi \sim Q/r$. We can estimate the size of the radius $r_*$ as the distance at which the non-linearity due to ${\cal O}_{n,k}$ become as important as the kinetic term. We then obtain 
\begin{equation}
r_* \sim \frac{1}{M_*} \left( \frac{\sqrt{s}}{M_*}\right)^{\frac{n-2}{n+4k-6}}.
\label{rsdy}
\end{equation}

Let us now reconsider the same process without solving any dynamical equation, but using only dimensional analysis. The two dimensionful parameters characterizing the process are the kinematic variable $\sqrt{s}$ and the coupling $G_\phi$. To illustrate the quantum nature of the scales, it is convenient to insert back the $\hbar$ factors, while working in units with $c=1$. In this way, we distinguish between mass ($M$) and length ($\ell$) scaling dimensions which, for the relevant quantities, are given by
\begin{equation} 
 [\hbar]=M \ell \, ,  ~~~~[{\cal{L}}]=M\ell^{-3}\, ~~~~ [\phi]=M^{\frac 12}\ell^{-\frac 12}\,  , ~~~~[\sqrt{s}]=M \, ,  ~~~~     [G_\phi]=M^{\frac{2-n}{2}}\ell^{\frac{n+4k-6}{2}}.
\, .
\end{equation}

Just with the help of dimensional analysis, we can construct some length scales out of the relevant dimensionful parameters. From $G_\phi$ and $\sqrt{s}$ we can determine the lengths
\begin{equation}
 L_*= G_\phi^{\frac{1}{n+2k-4}} \hbar^{\frac{n-2}{2(n+2k-4)}}\, ,
\end{equation}
\begin{equation}
 L=\frac{\hbar}{\sqrt{s}}\, .
\end{equation}
The length $L_*$ corresponds to the scale at which the
quantum effects of the  $\phi$ interaction become strong. The length $L$ 
 is the de Broglie wavelength of the source, which is the distance
at which the quantum properties of the localized waves are important. Both these scales incarnate quantum features of the process. In fact, they both vanish in the limit $\hbar \to 0$. 

Combining $G_\phi$ and $\sqrt{s}$, we can define another length
\begin{equation}
r_* = G_\phi^{\frac{2}{n+4k-6}}\sqrt{s}^{\frac{n-2}{n+4k-6}}=L_* \left( \frac{L_*}{L}\right)^{\frac{n-2}{n+4k-6}}.
\label{scall}
\end{equation}
This length persists in the limit $\hbar\to  0$: it is a classical length. It describes the distance at which the classical interaction of $\phi$ becomes strong.
When $L   \ll   L_*$, the classical
length $r_*$ is larger than any quantum length. Quantum effects become unessential
and the scattering process can be characterized in terms of a (semi)classical description. Note that the scale $r_*$ in eq.~(\ref{scall}), derived from dimensional arguments, parametrically coincides with the scale obtained in eq.~(\ref{rsdy}) from the dynamics, providing further evidence for the classical nature of the classicalon configuration.

The classicalization radius $r_*$  has the meaning of the distance at which a physical quantity associated with an observable becomes of order one in $M_*$-units. In particular,  this implies that the scattering of a probe particle off such a configuration becomes strong at such a distance, similarly to the deflection of photons in the gravitational field of a black hole near the Schwarzschild horizon. 
 In contrast to the solitonic case,  the $r_*$-radius does not necessarily coincide  with the maximal-energy localization region.

    \section{Electric field classicalized  by  topological charges}
 \label{sec3}
    
  In this section we would like to illustrate how the source of classicalons can also be topological charges.  As explained in the introduction, such  object cannot be used 
  for unitarizing the high-energy scattering amplitudes, since production of topological charges  
is exponentially suppressed in such processes.   Nevertheless, this example can serve as  an interesting illustration of how sharply localized topological charges lead to the creation of  classicalons
of much larger $r_*$-radius. 

 Our  example deals with correlators that
 would be naively UV-sensitive to highly localized charges.
 In this case an attempt to localize 
     charge at distances shorter than a certain length scale  $L_*$  results in its classicalization, 
 a turnover into a  classical object of size $r_*\gg L_*$  (in this respect classicalization is  related to the $r_*$-phenomenon discussed in 
ref.~\cite{dhk} so we borrow some examples from there).  
 
  Consider the following action:  
 \begin{equation}
\label{caction}
-\frac{F^2}{4}\, +\, L_*^4\left( \frac{F^2}{4} \right)^2 \, + \,  A_{\mu} \, j^{\mu} \, ,  
\end{equation}
where as usual $F^2 \, \equiv \, F_{\mu\nu}F^{\mu\nu}$. 
This action describes the interaction of an abelian gauge field to the current $j^{\mu}$. 
The gauge field also posseses a non-linear self-coupling described by the second term in eq.~(\ref{caction}). 
The crucial ingredient is that the current $j^{\mu}$ is topological, and thus is trivially conserved, 
$\partial_{\mu} \, j^{\mu} = 0$.  Thus,  $\int d^3x j^0\, \equiv  Q$ is a topological charge. 
  The precise nature of this topological charge is unimportant as long as it can be localized 
  within a finite region (sphere) of characteristic radius $L$. An explicit example
  will be constructed below. 

  The reason why we assume a topological charge is that, in this case, the 
  value of the charge is set by the localization width,  so that   $Q = L_*/L$, where the
  coefficient is  absorbed in  the normalization of $Q$. 

  Let us now show that any attempt of localizing $Q$ within a sphere $L \, \ll \, L_*$ will 
  result into the formation of a classical object of radius  $>> L_*$. 
   Consider the equation of motion
 \begin{equation}
   \partial^{\mu} \left[  F_{\mu\nu} \left(1 - \frac{L_*^4F^2}{2}\right)\right] \, = \, -j_{\nu}  \, , 
 \label{eqcharge}  
 \end{equation} 
    for a static charge $j_{\nu}  =  \delta_{\nu}^0 \, q(r)$ localized within a sphere of radius $L$. 
 We look for a static spherically symmetric field configuration $A_j = 0,  A_0 = \phi(r)$.  
Since $q(r)$ is localized within the radius $ L \, \ll \, L_*$, for $r \, \gg L$, the 
source can be replaced by $q(r) \, \rightarrow \, 4\pi \delta(r) Q$.   Equation~(\ref{eqcharge}) then becomes an algebraic cubic equation 
for $\partial_r\phi$, 
 \begin{equation}
   \partial_r\phi [1 + L_*^4(\partial_r\phi)^2)] \, = \, \frac{Q}{r^2} \, , 
 \label{eqcharge1}  
 \end{equation} 
which is easy to solve.  The important thing is the existence of the scale  $r_*\, = \, Q^{1/2} L_* = 
L_* (L_*/L)^{1/2}$.  This is the scale for which the value of the electric field saturates the cutoff 
$1/L_*^2$, and thus $r_*$ is the classicality radius of the charge $Q$.  We can see that 
for $r \, \gg \, r_*$,  the electric field is weak, $\partial_r\phi \sim Q/r^{-2}$, and hits the cutoff at $r=r_*$. 
For  $r \, <  r_*$, we have $\partial_r\phi \sim r^{-2/3}L_*^{-1}L^{-1/3}$.
 Therefore any attempt of localizing the charge within a radius $L$ creates a classical field configuration of much larger radius $r_*$.
 
 A consequence of this is that the scattering cross section of a probe on a localized charge 
grows at least as the geometric cross section, $\sigma \sim r_*^2\, = \, L_*^3/L$, for $ L \rightarrow 0$.   
  

 As an example,   we can resolve  $j^{\mu}$ by the topological current of a magnetic monopole.
  Then $Q$ will be proportional to the magnetic charge of the monopole.
  Consider a 't~Hooft-Polyakov monopole formed by a scalar field $\phi^a$,  with the 
  potential 
  \begin{equation}
  \lambda^2 (\phi^a\phi^a - v^2)^2\, ,
  \label{pot}
  \end{equation}
  where $a=1,2,3$ is
  an $SO(3)$ index.   The non-abelian field strength is $G_{\mu\nu}^a$.   The coupling 
  $A_{\mu} j^{\mu}$ up to a total derivative is generated by a coupling \cite{giaquantum}
  \begin{equation}
  L_* \,  F_{\mu\nu} (\phi^a G^a_{\alpha\beta}) \epsilon^{\mu\nu\alpha\beta}\, .
   \label{FF} 
  \end{equation} 
  Notice that, after plugging the expectation value of the scalar $\phi^a$, the term (\ref{FF}) 
  reduces to a theta-type mixing term between $F$ and the abelian subgroup of  $SO(3)$ 
  (call it $F^{M}$) under which the monopole has  a long-range magnetic field.  The magnetic charge of 
  the monopole under $F^M$  is $\mu\, = \, 1/g$, where $g$ is the gauge coupling of 
  $SO(3)$.   
  It is easy to see 
  that the magnetic charge under $F^{M}$, due to the above term, automatically becomes an electric charge under $F$ with 
  \begin{equation}
  Q \, = \, (L_*v)/g \, .
  \label{Qmonopole}
 \end{equation}
   (In fact, we can view this phenomenon as a generalization of Witten's mechanism for monopole 
  electric charge in the presence of theta-term\cite{witten}). 
 In the absence of (\ref{FF}), the magnetic charge of the monopole could be localized 
 at arbitrarily short distances by taking  $v \rightarrow \infty$ (keeping the gauge coupling $g$ fixed).   However,  the coupling (\ref{FF}) induces an electric 
 charge $Q$.   
   The crucial point is that the localization of this electric charge to arbitrary short distances is no longer possible.  Such attempt automatically classicalizes the electric charge. 
But in the presence of the coupling (\ref{FF})  the effective 
electric charge grows as $Q = vL_*$, and thus the electric charge classicalizes with the growing radius  $r_*  =  (L_*v)^{1/3} L_* $.  In other words, our gauge interaction plays a role similar to gravity. 
 
 \section{Classicalization of  Energy Sources } 
  
   Our second example concerns a scalar field with action, 
 \begin{equation}
\label{saction}
 (\partial_{\mu}\phi)^2 \left(\frac12 \, + \,  \frac{L_*^3}{4}~ \Box \, \phi \right) \, + \,  \phi  \, J .
\end{equation}   
   The idea is that this theory also classicalizes itself at distances below $L_*$. 
  We can read-off this result  from the solutions of this theory existing in the literature \cite{ddgv, andrei, nr, dhk}.
   In other words,  the guess  is that  any information stored in wave-packets of $\phi$ localized at the scale 
   $L \, \ll \, L_*$ (and readable by detectors sensitive to $\phi$-quanta) results into classical objects of size $r_* \sim L_* (L_*/L)^{1/3}$. 
   This is suggested by the fact that $\phi$ becomes classical at distances $r_*$ for any source 
   localized within a length $L$ and having a strength  $1/L$.  
   Consider a localized source at scale $L \ll L_*$.  At distances  $r \gg L$ such a source can be approximated by $4\pi \delta(r)/(M_*L)$.   The equation of motion 
   \begin{equation}
   \Box \, \phi + \, \frac{L_*^3}{2} \left[ (\Box \, \phi)^2-(\partial_{\mu}\partial_{\nu} \phi)^2 \right] \, = \, J \, 
   \end{equation}
   for this spherically symmetric ansatz simplifies to
   \begin{equation}
   \partial_j  \left[ \partial_j \phi(r) \,  - \, {x_j \over r^2} L_*^3 (\partial_r\phi)^2 \right] \, = \,-4\pi \delta(r) \, \frac{L_*}{L} ,
  \label{sphere}
  \end{equation}
  which can be easily integrated and rewritten as a quadratic  algebraic equation for 
  $\partial_r\phi(r)$, 
  \begin{equation}
  \partial_r\phi \, - \, \frac{L_*^3 (\partial_r\phi)^2}{r} \, +\,   \frac{L_*}{  L  r^2}\, =\, 0 \, .
\label{algebra}
\end{equation}    
The exact solution of this equation\cite{nr}
\begin{equation}
\partial_r\phi \, = \, {r \over 2L_*^3} \left( 1\pm \sqrt{1+\frac{4r_*^3}{r^3}}\right)
 \label{exact}
\end{equation}
shows that  at distances  $r \, \gg \, r_* \equiv \,  L_* (L_*/L)^{1/3}$, 
we have $\phi \, \sim \, L_*/(Lr)$,  whereas for $r \ll r_*$  we have $\phi \, \sim \, \sqrt{r/L}\, L_*^{-1}$. 
Thus, $r_*$ plays a role similar to the Schwarzschild radius and is the scale  at which $\phi$ classicalizes. The analogous phenomenon in massive gravity is known under the name of Vainshtein effect~\cite{vainshtein}. 
   
   Now it is clear that the radius $r_*$ is given by the distance at which the first term in eq.~(\ref{algebra}) becomes comparable to the second one.  Starting at large distances, where  $\phi = (M_*L r)^{-1}$ and comparing the two terms, we find 
 $r_* = L_* (L_*/L)^{1/3}$.  So we see that, although the interaction of $\phi$ looks horribly 
 bad in the UV, any integration over sources localized at the scale $L$ must be suppressed, because such sources have typical size of order $r_*$, and thus 
 should be classicalized.
 
 Indeed,  consider the scattering of two sharply localized wave-packets prepared either out of 
 $\phi$-quanta or any other particles that source $\phi$.  
 For instance, imagine that $\phi$ is sourced by the trace of the energy momentum tensor 
 of some fermions $\psi$, 
 \begin{equation}
 L_* \phi T_{\mu}^{\mu}(\psi)\, .
 \label{tsource}
 \end{equation} 
 Perturbatively, the exchanges of the $\phi$-quanta violate unitarity for the 
 scattering of $\psi$ particles with momentum transfer above the scale $M_*$. 
 Non-perturbatively, however,  as long as the impact parameter 
 is within the $r_*$ radius of an effective source, the scattering must become dominated 
 by classical configurations of the $\phi$ field.  This intuitive argument  can be made more precise 
 by the following consideration.    Relativistic pulses that 
 come within distance $L$ and experience momentum transfer $1/L$, on time scales $L$ can be approximated as a static  source  of mass $1/L$ and size $L$. The $\phi$ field created by such a source can be found from eq.~(\ref{algebra}) and has an associated $r_*$-radius given by  $r_* = L_*(L_*/L)^{1/3}$, which exceeds 
 $L_*$ for  $L < L_*$.   Thus, the scattering amplitudes classicalize above $M_*$. 
 
  The same must be true for any virtual quanta that source
 $\phi$. Integration over sources with $r_*$ radius exceeding $L_*$, implies integration over 
{\it classical}  objects and must be exponentially suppressed. 
 So the basic ingredient for classicalization by $\phi$ is the existence of interactions 
  that lead to the $r_*$-phenomenon, which requires the existence of appropriate non-renormalizable operators.   Let us now discuss some speculative applications to the  Standard Model physics.

\section{UV-Completion of the Standard Model by Classicalization?}. 

In order to classicalize the Standard Model the three logical possibilities are:

 1) In the absence of Higgs, the classicalizer is an additional field \cite{giacesar}.  Such a case is effectively in the same class as classicalization by gravity. 
   
  2) Classicalizers are the longitudinal $W$-bosons.   

 3) The classicalizer  is the Higgs.

 \subsection{UV-Completion  of the Standard Model  without Higgs by Classicalization}
 
   Let us first discuss the non-Wilsonian UV-completion of the Higgsless Standard Model 
 by  classicalization.
  For this we either need to introduce an additional classicalizing  field, such as gravity, or 
classicalize the interactions of longitudinal $W$-bosons  without introducing new physics. 
 Let us turn to the second option which is the most dramatic one. 
  We shall first consider  a prototype model~\cite{dhk}, which illustrates how the interactions of longitudinal 
  vector field exhibit the $r_*$-phenomenon and thus classicalize in the absence of the Higgs.   Consider a  massive vector field $W_{\mu}$
  (for the moment we shall limit ourselves to a single self-interacting vector field). At the linear 
  level the dynamics of this field is described by the well-known Proca  Lagrangian 
  \begin{equation}
  {\cal L} \, =\, -{1 \over 4} F_{\mu\nu}F^{\mu\nu} \, + \, {1\over 2} m_W^2\, W_{\mu}W^{\mu}  \,  .
  \label{proca}
  \end{equation}
  The $W_{\mu}$ field propagates $3$ degrees of freedom which, without loss of generality, can be written as 
  \begin{equation}
  W_{\mu} \, = \, \tilde{W}_{\mu} \, + \, {1 \over m_W} \partial_{\mu} \phi \, . 
  \label{Stuckelberg}
  \end{equation}
 Here $\tilde{W}_{\mu}$ describes the transverse polarizations of the $W$-boson and $\phi$ is a canonically-normalized St\"uckelberg field describing its longitudinal polarization. 
  This field gets its kinetic term from the mass term of the gauge-field in eq.~(\ref{proca}). This is the source of the 
  problem. In the presence of self-interactions, perturbative unitarity 
is violated above the scale set by $m_W$.  In the standard approach one therefore integrates in the 
Higgs boson in order to restore unitarity.  

 We now wish to reconsider this violation of unitarity in the light of classicalization. 
 We wish to show that  self-interactions of the longitudinal Proca field classicalize exactly above the 
 scale where naive perturbative unitarity is violated.   As shown in ref.~\cite{dhk}, the longitudinal Proca field
 exhibits the $r_*$-phenomenon, and we will discuss here how this is essential for classicalization. Let us add a simple self-interaction to the Proca Lagrangian 
   \begin{equation}
  {\cal L} \, =\,  -{1\over 4} F_{\mu\nu}F^{\mu\nu} \, + \, {1\over 2} m_W^2\, W_{\mu}W^{\mu}  \,   + \, 
  {g^4 \over 4} \,  \, (W_{\mu}W^{\mu})^2 \,  - \, g W_{\mu} j^{\mu}
  \label{procaint}
  \end{equation}
where $j^{\mu}$ is an external source and $g$ is a coupling constant. The new interaction can be viewed as the first term of an operator expansion. Note that, due to the non-zero mass, conservation of  $j^{\mu}$ is no longer mandatory.  Instead, the equation of motion  implies the following 
constraint, 
   \begin{equation}
   \partial^{\mu} \left[W_{\mu} \, \left(m_W^2  \,   + \, 
  g^4  \,W_{\nu}W^{\nu}\right) \right] \,  =   \, g\partial^{\mu} j_{\mu} \, .
  \label{constraint}
  \end{equation}
The quartic  interaction generates the following self-coupling for the longitudinal 
(St\"uckelberg) component,   
 \begin{equation}
\label{nambu}
{g^4 \over 4  m_W^4} (\partial_{\mu}\phi\partial^{\mu}\phi)^2 \, .
\end{equation}  
Moreover, an exchange of longitudinal $W$-bosons induce the interaction 
\begin{equation}
{g^2\partial^{\mu}j_{\mu} \partial^{\nu}j_{\nu}  \over m^2_W (m_W^2 + \Box )} \, .
\label{source2}
\end{equation}
The operators in eqs.~(\ref{nambu}) and (\ref{source2}) violate
 perturbative unitarity at distances shorter than $L_* \equiv  g/m_W$. 
However, the sources classicalize at such a distance.   In order to see
 this, consider a $\phi$-field created by a point-like source
 of strength $1/L$ of the form $\partial^{\mu}j_{\mu} = \delta(r) /L$.  To see the effect of this source we can decouple the transverse gauge field, by taking the limit  $g \rightarrow 0$ with $L_*$ fixed.  
Equation~(\ref{constraint})  then becomes the equation of motion for  $\phi$, which takes the form   (\ref{eqcharge1}) and which we have solved previously. Its solution  indicates that the 
sources classicalize at the distance $r_* =  L_* \sqrt {L_*/L}$.  This classicalization is entirely due to 
interactions of longitudinal $W$-boson and is unrelated to the presence of the Higgs particle. 
To see this, we can explicitly rewrite the Lagrangian  (\ref{procaint}) as the  Higgs-decoupling limit of the theory with  the Higgs.  For this we can first integrate in the Higgs field by 
expressing $\phi$ as the phase of a complex scalar $H \, = \, \rho(x) e^{i{\phi(x) \over v}}/\sqrt{2}$ with Lagrangian  
   \begin{equation}
{\cal L} \, =\,  -{1\over 4} F_{\mu\nu}F^{\mu\nu} \, +  D_{\mu} H^* D^{\mu} H\, + \, 
  {1\over v^4} \,  (D_{\mu} H^* D^{\mu} H)^2 \,  -  {\lambda^2 \over 2} \, \left(H^*H \, - \, \frac{v^2}{2}\right)^2 \, ,
  \label{Higgsaction}
  \end{equation}
  where $D_{\mu} \, = \, \partial_{\mu} \, - \, i g\, W_{\mu}$.  This theory contains two particle mass scales, the gauge boson mass $m_W = gv$ and the Higgs mass $m_H \, = \, \lambda v$. Perturbative unitarity is violated at the scale  $M_* \, = \, v \, = \, m_W/g$.    We can keep 
  the scales $m_W$ and $M_*$ fixed, but decouple the Higgs by taking the limit 
  $\lambda \, \rightarrow \, \infty$.  In this limit the Lagrangian~(\ref{Higgsaction})
  reduces to (\ref{procaint}).

 Another interesting classicalizing interaction that we can add to the Proca Lagrangian (\ref{proca}) has the form
 \begin{equation}
 g^3 \, (\partial_{\mu} W^{\mu})   W_{\nu} W^{\nu} \, .
 \label{wint}
 \end{equation}
 After taking into account the decomposition (\ref{Stuckelberg}), the above interaction generates the following interaction for the longitudinal component
  \begin{equation}
\label{nambu1}
{g^3 \over m_W^3} (\partial_{\mu}\phi)^2 \Box \, \phi.
\end{equation}   
 Notice that this interaction exactly reproduces the second term in the scalar example of eq.~(\ref{saction}), 
 with $L_* \, = \, g/m_W$.   In fact, by taking the decoupling limit  $g \, \rightarrow \, 0$ with   $g/m_W$  fixed, the Proca theory  with interaction (\ref{wint})  exactly reproduces  the theory of eq.~(\ref{saction}) plus a decoupled massless gauge field $\tilde{W}_{\mu}$. 
 
 But as we have seen the interaction in eq.~(\ref{nambu1}) has the property 
 to classicalize at $L_*$.  Thus, although the scattering of longitudinal $W$-bosons violates perturbative unitarity  at distances shorter than  $g/m_W$, non-perturbative arguments tell us that  this is precisely the barrier below which the interaction classicalizes itself and scattering are dominated by creation of extended classical objects.   
  
   To discuss the realistic Standard Model we have to generalize the above classicalization mechanism  to non-abelian theories.   In the non-abelian case, the gauge field mass term automatically contains the self-interactions that violate perturbative unitarity. These come from the following 
   terms in the Lagrangian, 
   \begin{equation}
   [D_{\mu} U(x)]^{\dagger a}  [D_{\mu} U(x)]_{a} \, ,
   \label{nonabelian}
   \end{equation}
where $U(x)_a, ~(a=1,2)$ represents  an $SU(2)$ St\"uckelberg field  that is  obtained by the action of local $SU(2)$-transformation upon a constant doublet $(0, v)$.   The interactions of the St\"uckelberg fields
violate perturbative unitarity at the scale $M_* = v$. The question is whether the same interaction 
is able to classicalize the theory above the scale $v$ or whether additional classicalizing interactions of the form, 
  \begin{equation}
 \left( [D_{\mu} U(x)]^{\dagger a}  [D_{\mu} U(x)]_{a}\right)^2 \, + \,  L_*^4 \, \, U(x)^{\dagger b} [D^{\nu} D_{\nu} U(x)]_{b} \, [D_{\mu} U(x)]^{\dagger a}  [D_{\mu} U(x)]_{a} \, 
   +...
   \label{clas2}
   \end{equation}
 are required.  We shall now investigate this question.  
 
 \section{Classicalization of Nambu--Goldstone Bosons} 
 
     We shall  first study the role of  higher-dimensional operators in the classicalization of Goldstone fields.   Consider an abelian Goldstone model with a complex scalar field  
$H \, =e^{i\theta} \rho(x) /\sqrt{2}$ and with the 
following action,  
   \begin{equation}
  \partial_{\mu} H^* \partial^{\mu} H\,   -  {\lambda^2 \over 2} \, \left(H^*H \, - \, 
  \frac{v^2}{2} \right)^2 \,  -  (\partial_{\mu}\theta)J^{\mu}\,,
  \label{ng0}
  \end{equation}
where $\theta \, \equiv \, \phi / v$ is a Goldstone angular degree of freedom, and $\rho(x)$ 
is a radial (Higgs) degree of freedom with mass $m_H \, = \, \lambda v$.   Here $J^{\mu}$ is an external current, which for example, can be taken as the axial fermionic current 
$J^{\mu} \, = \, \bar{\psi} \gamma_5\gamma^{\mu} \psi$.   Rewriting the action in terms of the Higgs and Goldstone degrees of freedom, we obtain 
   \begin{equation}
  {1 \over 2} (\partial_{\mu} \rho)^2  \, +\, \frac{\rho^2}{2} (\partial_{\mu}\theta)^2
 \, -\,  {\lambda^2 \over 8} \, (\rho^2 \, - \, v)^2\, -\,  (\partial_{\mu}\theta)J^{\mu}\, .
  \label{ng1}
  \end{equation}
 At distances $\gg m_{H}^{-1}$, we can integrate out the Higgs degree of freedom by replacing it with its expectation value, and write down an effective theory of a Goldstone field 
   \begin{equation}
\frac   {v^2}{2} (\partial_{\mu}\theta)^2   -  (\partial_{\mu}\theta)J^{\mu}  \,  + \,  {\rm higher}{\mbox -}{\rm dimensional~operators} \,.
  \label{nambugold}
  \end{equation}
Although the scale that suppresses the Goldstone interactions is 
$v$, the higher-dimensional operators will be suppressed by powers of $m_H$, which 
in a weakly coupled theory ($\lambda \ll 1$) is below $v$.   Let us ignore these operators for a moment. Then, the Lagrangian in eq.~(\ref{nambugold}) describes a theory with cutoff $L_* = v^{-1}$. 
Let us study the classicalization phenomenon in such a theory.  Proceeding as in the previous 
examples, let us prepare a spherical source localized over distance $L$ and strength $1/L$, which 
for $r \gg L$ can be approximated by  $\partial_{j} J^{j}  = 4\pi \delta(r)/L$.  According to the equation of motion, 
  \begin{equation}
  \Box \, \theta(x) \, =\, \frac{\partial_\mu J^{\mu}}{v^2}  \,, 
  \label{ngequ}
  \end{equation}
such a source results into a classical Goldstone field, 
 \begin{equation}
\partial_r \theta(r) \, = \, \frac{L_*^2}{Lr^2} \, .
  \label{ngfield}
  \end{equation}
 In order 
to study the classicalization phenomenon, we are interested in sources localized at 
$L \, \ll \, L_*$.   For classicalization to happen, two conditions are necessary.  First the source 
should be able to provide a value of the Goldstone gradient of order  $1/L_*$,  that is 
$\partial_r\theta(r) \, \sim \,  v$. Seemingly this is the case for the solution (\ref{ngfield}) at sufficiently small $r$, but this is just an illusion created by the low-energy approximation.   The culprit is the Higgs field,  which prevents the growth of the  Goldstone-gradient 
beyond the scale $m_H$.  To see this, we can examine the back-reaction of the 
solution (\ref{ngfield}) on $\rho$ by inspecting the equation of motion 
for  $\rho$ evaluated on the background solution (\ref{ngfield}), 
   \begin{equation}
  \left[ \Box \, +\,
  (\partial_r\theta)^2 \, - \,  
  \frac{\lambda^2 v^2}{2}\, + \, \frac{\lambda^2}{2} \rho^2      \right] \rho = \, 0 \, .
  \label{higgsstability}
  \end{equation}
  Replacing $\partial_r \theta$ with the background solution (\ref{ngfield}),  it is clear that the tachyonic mass for 
$\rho$ changes sign and becomes positive for $r \, < \, v^{-1} /\sqrt{Lm_H}$.    
   As a result, in the weakly-coupled Goldstone-Higgs system, classicalization  does not happen. 
  
   Now it is clear that, in order for classicalization to take place, we have to take the limit
   $\lambda \rightarrow \infty$, keeping $v$ fixed and, at the same time, we have to supplement 
 the theory by a  higher-dimensional operator  $L_*^4 (\partial_\mu\theta\partial^{\mu}\theta)^2$. 
 This is precisely the operator that is obtained by integrating out the Higgs field, but 
 for classicalization to happen we have to keep its coefficient fixed  $L_* = v^{-1}$ even in the limit 
 of infinitely-heavy Higgs.    
  This  would guarantee the turn-over  of the solution at the scale $r_*$. 
 
  \subsection{Non-abelian Goldstone theory} 
  
  In order to reproduce the Standard Model, we now consider the non-abelian generalization of the previous model to an $SU(2)$ global symmetry,  
 \begin{equation}
 \partial_{\mu} H^{a*} \partial^{\mu} H_a\,   -  {\lambda^2 \over 2} \, \left(H^{a*}H_a \, - \, \frac{v^2}{2}\right)^2 \,  -  H_a {\bar \psi}^a \psi  + \, i \, {\bar \psi}^a\gamma_{\mu} \partial^{\mu} \psi_a  \, + ...\, .
  \label{ng}
  \end{equation}
Here $H_a\, (a=1,2)$ is an $SU(2)$-doublet scalar field, and in order to provide for an external test-source we have introduced $SU(2)$-doublet ($\psi_a$) and $SU(2)$-singlet ($\psi$) fermions. 
   We shall now  represent the doublet  field in terms of the radial and Goldstone degrees of freedom,  
$H_a \, = \,  U_a(x) \,\rho(x)/\sqrt{2} \, = \, 
 ( cos\theta e^{i\alpha}, \, -sin\theta e^{-i\beta} )\rho/\sqrt{2}$, where 
$\theta, \alpha$, and $\beta$ are the three Goldstone fields of the spontaneously broken global 
$SU(2)$  group.     Now we wish to decouple the Higgs particle by taking the limit
$\lambda = \infty$ with $v $ fixed.  In the remaining sigma-model the non-linear self-interactions of the Goldstone triplet is described by the following invariant 
\begin{equation}
 I \, \equiv \, v^2\, \partial^{\mu} U(x)^{\dagger}  \partial_{\mu} U(x) \, = \, 
  v^2 \left[ (\partial_{\mu} \theta)^2  + \cos^2 \theta  (\partial_{\mu} \alpha)^2   +   \sin^2\theta (\partial_{\mu} \beta)^2  \right ] \, . 
  \label{invariant}
  \end{equation} 
 At a first glance, these non-linear self-interactions have the right structure for classicalizing 
 the localized sources and creating the scale $r_*$.   However,  although the classical configurations are produced, the $1/r^2$-growth of the Goldstone gradients for $r \rightarrow 0$  cannot be regulated by non-linearities for any localized  source.   
  To illustrate this, let us take an external fermionic current to be composed out of fermions with only  first non-zero $SU(2)$-component,
$J_{\mu} = \bar{\psi}^1\gamma_{\mu} \psi_{1}$, 
 \begin{eqnarray}
&{\cal L}=I +i\left[ (\partial_\mu U^{\dagger})U -U^{\dagger}(\partial_\mu U)\right] J^\mu = 
\nonumber \\ &
 v^2 (\partial_{\mu} \theta)^2  + v^2 \cos^2 \theta \left[ (\partial_{\mu} \alpha)^2 \, + \,  2(\partial_{\mu} \alpha) \frac{J^{\mu}}{v^2} \right]  +   v^2 \sin^2\theta \left[ (\partial_{\mu} \beta)^2   \, - \,2(\partial_{\mu} \beta) \frac{J^{\mu}}{v^2} \right] \, .
   \label{ngsigma}
  \end{eqnarray}
 The equations of motion are 
   \begin{equation}
   \partial^{\mu} \left [ cos^2 \theta  \left(\partial_{\mu} \alpha \, + \,   \frac{J_{\mu}}{v^2} \right) \right ]\, = \, 0  \, ,
   \end{equation}  
  \begin{equation}
   \partial^{\mu} \left [  sin^2\theta \left(\partial_{\mu} \beta  \, - \, \frac{J_{\mu}}{v^2} \right )\right] =  0 \, ,
   \end{equation}
   \begin{equation}
   \Box \, \theta +\frac{\sin 2\theta}{2} \left[ (\partial_{\mu} \alpha)^2 - (\partial_{\mu} \beta)^2 +2\partial_\mu (\alpha +\beta ) \frac{J^\mu}{v^2} \right]
=  0 \, .
   \label{eqsforsigma1}
  \end{equation}
 These equations of motion are trivially satisfied by  
 $\partial_{\mu} \alpha =  - \, \partial_{\mu} \beta = - J_{\mu}$,  and by  $\theta$  being  an arbitrary 
 plane-wave satisfying the massless wave equation $\Box \, \theta \, = \, 0$.  In particular, for an arbitrary 
 localized spherical source,  $\partial_{j} J^{j}  = 4\pi \delta(r)/L$, the system allows the Goldstone gradients to grow unbounded at short distances.  Thus,  although classicalization 
 takes place, the system is unable to regulate the classicalized solutions at short scales solely by 
 Goldstone self-interactions, and requires introduction of  higher powers of the invariant  $I$.
 For example, consider adding the square of this invariant  to the original action (\ref{ngsigma}), so that the resulting Lagrangian  becomes 
 \begin{equation}
{\cal L} \, = \,    I \, + \, \frac{I^2}{2v^4}\, +  2 (\cos^2 \theta \, \partial_{\mu} \alpha  -  \sin^2\theta \, \partial_{\mu} \beta) J^{\mu} \, .
   \label{regulators}
   \end{equation}
 The equations of motion now become,
   \begin{equation}
   \partial^{\mu} \left\{  \cos^2 \theta  \left[ \left(1 +  \frac{I}{v^4}\right)\partial_{\mu} \alpha \, + \,   \frac{J_{\mu}}{v^2}\right] \right\} \, = \, 0 \, ,
   \end{equation}  
  \begin{equation}
     \partial^{\mu} \left\{  \sin^2 \theta  \left[ \left(1 +  \frac{I}{v^4}\right)\partial_{\mu} \beta \, - \,   \frac{J_{\mu}}{v^2}\right] \right\} \, = \, 0 \, .
   \end{equation}     
   \begin{equation}
  \partial^\mu \left[  \left(1 +  \frac{I}{v^4}\right) \partial_{\mu}\theta \right]
  +\frac{\sin 2\theta}{2} \left\{ \left(1 +  \frac{I}{v^4}\right) \left[ (\partial_{\mu} \alpha)^2 - (\partial_{\mu} \beta)^2 \right]+2\partial_\mu (\alpha +\beta ) \frac{J^\mu}{v^2} \right\} =0 \, . 
  \end{equation}
A solution to these equations is found for constant $\theta$ and $\partial_r \alpha  = - \partial_r \beta = f(r)$ satisfying the algebraic equation
 \begin{equation}
  f(r)- L_*^2f(r)^3 +  \frac{L_*^2}{Lr^2} =0   \, ,  
  \end{equation}
  where $L_* = 1/v$.
 As we have already shown in sect.~\ref{sec3}, this equation exhibits classicalization at the scale $r_* =  L_*(L_*/L)^{1/2}$.  This result can be easily generalized to the case in which the Lagrangian is an arbitrary function of the invariant $I$, 
   \begin{equation}
{\cal L} \, = \,   F(I) \,  +  2 (\cos^2 \theta \, \partial_{\mu} \alpha  -  \sin^2\theta \, \partial_{\mu} \beta) J^{\mu} \, .  
\label{FI}
\end{equation} 
 For a spherically-symmetric localized source, the  equations of motion are solved by the same ansatz ($\partial_r \alpha  = - \partial_r \beta = f(r) $ and $\theta$ constant), with the 
 function $f(r)$ now satisfying the equation
 \begin{equation}
\left  [{d F(I) \over dI} \right ]_{I=-\frac{f^2}{L_*^2}} \, f(r) \, = \, - \frac{L_*^2}{Lr^2}  \, .
 \label{arbitrary}
 \end{equation}
For any regular function $F(I)$ expandable in powers of $I/v^4$, the above solution classicalizes at the scale  
 $r_* \sim  L_*(L_*/L)^{1/2}$.  
 The reason is that the scale $r_*$ is defined as the 
 distance at which non-linearities in $F(I)$ become important and, for a generic function $F$, this happens when $I(r) \sim   v^{4}$.
 For instance, for a (DBI-type) function,  $F(I)  = L_*^{-4} \sqrt{1 \, + \, L_*^4 I}$, the solution is 
 \begin{equation}
 f(r) \, = \,  {1 \over  L_*}\left( 1 \, + \, \frac{L^2r^4}{4L_*^6} \right)^{-1/2} \, .
 \label{root}
 \end{equation}
 
  In each order in fields, the addition of higher-derivative  invariants is not changing the value of the scale $r_*$, but only the behavior of the solution  at the short distances. Thus, the classicalization scale is rather insensitive  to higher derivatives. 
 
     If the above classicalization of the Goldstone modes works,  the recipe for constructing the classicalizing Higgless  standard model is the following.   Supplement the action of the Standard Model with the higher-order invariants of the form,  
 \begin{equation}
   v^{-4}(D_{\mu} H^{\dagger }  D_{\mu} H)^2 \, + \, ...\, .
   \label{higgsinv}
   \end{equation}
Then take the limit $\lambda \to \infty$,  keeping $v$ fixed.  In this limit  the radial Higgs  mode $\rho$ decouples, converting  the Higgs doublet into a St\"uckelberg doublet 
$H_a \rightarrow   U_a v/\sqrt{2}$. The Standard Model Lagrangian then becomes 
 \begin{equation}
   \frac{v^2}{2}(D_{\mu} U^{\dagger}  D_{\mu} U) \, + \,\frac14  (D_{\mu} U^{\dagger}  D_{\mu} U)^2 \, + \, ...\, .
  \label{higgsless}
 \end{equation}
In this way, we are left with a manifestly gauge invariant theory 
in which  the Higgs doublet is replaced by the St\"uckelberg doublet. The propagating degrees of freedom  of this theory are exactly the same as in the Standard Model without the Higgs. 
 Such a theory violates perturbative unitarity above the scale $v$, but in deep UV unitarity is restored by classicalization.  The theory above $v$ is no longer a Wilsonian quantum field theory, but rather a 
 theory of classicalized extended objects.   In other words, classicalization is the response of the  
 theory to the lack of Wilsonian UV completion. 
 
  In order  to see how the role of the Higgs is replaced by classicalization, notice that the 
  classicalizing higher-dimensional operators in eq.~(\ref{higgsless}) are of precisely the same form 
which  is obtained by integrating out the Higgs particle.   Indeed, ignoring fermions, the Higgs-dependent 
part of the Standard Model in our parameterization is
  \begin{equation}
 \frac12 (\partial_{\mu} \rho)^2 \, + \,  \frac{\rho^2}{2}(D_{\mu} U^{\dagger}  D_{\mu} U) \, - \, {\lambda^2 \over 8} (\rho^2 - v^2)^2 \, .
   \label{higgspart}
   \end{equation}
Integrating out the Higgs through its equation of motion,  which at low energies becomes an algebraic constraint
  \begin{equation}
  \rho^2 \, = \, v^2 +\frac{2}{\lambda^2} (D_{\mu} U^{\dagger }  D_{\mu} U) \, ,
   \label{higgseq}
   \end{equation}
we obtain the following effective theory
 \begin{equation}
   \frac{v^2}{2}(D_{\mu} U^{\dagger}  D_{\mu} U) \, + \, {1\over {2\lambda^2}}  (D_{\mu} U^{\dagger}  D_{\mu} U)^2 \,.
  \label{higgseff}
 \end{equation}
Notice that the second term exactly coincides with the classicalizing interaction
in eq.~(\ref{higgsless}), with the difference that it is suppressed by $\lambda^2$, and thus in the ordinary Standard Model 
would disappear in the limit of infinite Higgs mass.  Classicalization thus corresponds to the situation in which, although the Higgs has been eliminated, the would-be Higgs-mediated interaction is maintained.

\subsection{Phenomenology of  Higgless Classicalization}

  Let us discuss some phenomenological aspects of classicalization as seen from the low-energy  effective field theory point of view.   For this  let  us reiterate some key points of the classicalization phenomenon.   We shall work in the approximation in which there is a substantial hierarchy 
between the mass of the classicalizer field $\phi$ and the the classicalization scale $M_*$, so that 
for the energies of interest the classicalizer can be treated as massless. 

  In order to make some phenomenological predictions, we need to understand what are the UV-sensitive 
  and UV-insensitive aspects of the classicalization phenomenon. 
 We first wish to stress again that classicalization of sources with $1/L \, \gg \, M_*$ is predominantly a long distance phenomenon, largely insensitive to the structure of the theory 
 at distances $\sim \, L_*$.  The inevitability of classicalization of such sources can be understood already at the linearized level, as a consequence of the Gauss's law.  Let us take a source $J(r)$
localized  within a sphere of  radius  $r_* \, \sim \, L_*\sqrt{L_*/L}$,  and having an integrated value 
 $L_*  \int_0^{r_*} \, r^2 dr \, J(r) \, = \,  L_*/L$.
 According to the linearized equation 
 \begin{equation}
 \Box \, \phi \, +...\, = \, J(r) L_* \, , 
 \label{gauss2}
 \end{equation}
such a source is producing a Gaussian flux of the gradient $\partial_r \phi  \, \sim \,  L_*/(r^2L)$ that 
 approaches  values of order $M_*^2$,  for $r \rightarrow  r_*$.  When we are  approaching the source 
 from large distances,  the  effect of the non-linearities  is fully under control and 
 we can be sure that the quantity $L_*^2\partial_r\phi$ becomes of order one for distances 
 at which the effect of higher-dimensional operators is not yet dominant.  As a result, we are convinced that such a source produces a classical configuration of size $r_*$.  For example, a test source particle interacting with the gradient of $\phi$ will scatter at  such a configuration with a geometric cross-section $\sigma \, \sim \, r_*^2$.  

 Such a source is a highly classical state, and its decay into any two-particle quantum state is exponentially  suppressed.   This result indicates that the two-to-two scattering amplitude
 with momentum transfer $\sim 1/L  \gg M_*$ producing such  classicalized sources  
  must be exponentially suppressed.  
  
  
   As an example consider the effect of clasicallization  in a two-to-two scattering process
  mediated by a four-derivative interaction 
  \begin{equation}
  L_*^4 ( \partial_{\mu}\phi \partial_{\mu} \phi)^2 \, .
  \label{quartic}
  \end{equation}
  Such interaction models the scattering of longitudinal $W$-bosons in the Higgless 
  Standard Model, with $L_* = v^{-1}$.  Perturbatively, the amplitude of this process grows as 
  \begin{equation}
   A(s) \,  \sim  \frac{s^2}{v^4} \, .
\label{amplitude1}
   \end{equation} 
 However, classicalization shuts-off the growth for $\sqrt{s} \, > \, v$.  
In fact, for $\sqrt{s} \, \gg \,  v$, the  amplitude of the process $WW \rightarrow WW$ is completely dominated by the low momentum transfer of order\footnote{Notice that the appearance of  $(\sqrt{s})^{1/3}$ as opposed to $\sqrt{s}$ in eq.~(\ref{r1/3}) is due to the fact that 
$\phi$ is self-sourced not by  the two-derivative operator $(\partial\phi)^2$, but rather by a four-derivative one which, for the same $\sqrt{s}$, produces a shorter $r_*$-radius. This result is obtained from eq.~(\ref{rsdy}) for $n=4$ and $k=2$.}
\begin{equation}
r_*(s)^{-1} \, =\, v \left( \frac{v}{\sqrt{s}}\right)^{1\over 3} \, , 
\label{r1/3}
\end{equation}
and therefore
\begin{equation}
   A(s) \,  \sim [v r_*(s)]^{-4} \, \sim \, \left ({v \over \sqrt{s}}\right )^{4 \over 3} \, . 
\label{amplitude}
   \end{equation} 
 The reason for this suppression is  that  events with high momentum transfer
 ($\sim \sqrt{s}$) because of classicalization go 
 through the formation of an intermediate classical state of size $r_*(s)$ and mass
 $\sim \sqrt{s}$, whose decay probability into a two-particle quantum state is exponentially suppressed at least by the following factor,
 \begin{equation}
  \exp \left[{-\sqrt{s} r_*(s)}\right] \, \sim \,  \exp \left[{- \left ({\sqrt{s}\over v} \right )^{4 \over 3}}\right] \, .
  \label{exponent}
  \end{equation}
 Such a classical state will instead decay into many particles, with the decay being dominated by low-momentum states. Of course, the total inclusive cross section of the $WW$-scattering process 
increases as the geometric cross section 
\begin{equation}
\sigma \, \sim \, r_*(s)^2 \, \sim  \,  v^{-2} 
 \left ({\sqrt{s} \over v} \right ) ^{2 \over 3}.
\end{equation}

One may wonder how sensitive are the above features to the possible higher-dimensional operators that can source  $\phi$.  As previously discussed, they are not.   
For example, imagine that we add to the classicalizing  interaction other  terms that are higher-order in derivatives, say,  
\begin{equation}
\partial^k (\partial\phi)^{2n+2} \, . 
\end{equation}
In the scattering process each term individually self-sources $\phi$ and results 
in the production of a classical object of size 
\begin{equation}
r_*\, = \,  v^{-1} 
 \left ({\sqrt{s} \over v} \right ) ^{n+ 1/2 \over 3n + k +3/2}.
\label{rforn}
\end{equation} 
This  largest  $r_*$ corresponds to $k=0$. 
Thus, at high $s$, the scattering is completely dominated by the leading term, and higher-derivative 
terms are irrelevant. The 
insensitivity of the scattering amplitudes to the higher invariants for large values of $s$, is the key property of the classicalization phenomenon. 
On the other hand, the higher invariants will play an important role for $s \sim v^2$, where the dynamics becomes model dependent.
 
  The stability of the classical states is not an essential ingredient for the classicalization phenomenon, as long as the lifetime $\tau$ of the classical states of size $r_*$ is 
 such that  $\tau \simgt r_*$.  This condition should hold in any ghost-free
 theory with positive energy sources, and it can be translated into a restriction on the types of 
 higher-dimensional operators. 
 
   From the point of view of LHC phenomenology,  it is important to understand the dynamics of the
   states around $L_*$. 
   This dynamics is defined by the cross-over phenomenon according  to which any consistent 
   theory that classicalizes above the energy $1/L_*$ must contain quantum resonances with masses around this scale 
\cite{previous}.  The existence of such resonances inevitably follows from 
   the quantum-to-classical transition.   They correspond to the quantum states obtained from the classical configurations in the limit $r_* \rightarrow L_*$.  The key point is that the lifetime  of a classical configuration must be parametrically larger   
 than its characteristic size, $\tau \, >\, r_*$.  This is a defining property of a classical state and   
it should persist for any $r_* \, > \, L_*$.   Thus, classicality of objects with size $r_*$ implies the existence of quantum resonances with mass $L_*^{-1}$.  To be more precise, classicalization implies the existence of a tower of resonances that starts at $M_*$, with states becoming more and more classical as mass increases.  The heavier the resonance, the more classical it is, and more insensitive its properties are to higher-order invariants in the action.   
 
  Thus,  classicalization of the Higgless Standard Model leads us to the following 
self-consistent picture.   We start from 
the Standard Model and decouple the Higgs by taking the limit  $\lambda \to \infty$ with $v= M_*$ fixed. Despite the absence of the Higgs, the high energy  ($1/L \, \gg v$) scattering is now unitarized by classicalization. 
This fact on its own is insensitive to the structure of the higher-dimensional operators, since it  comes entirely from the long-distance effects of the localized sources.  On the other hand, 
the  precise structure of the higher-dimensional operators is important for defining the short-distance properties of the classicalized objects, and thus, for defining the dynamics of quantum resonances 
at the scale $M_*$.  Thus, higher-dimensional operators are crucial in setting the spectrum 
of states in the vicinity of $M_*$,  but quickly become irrelevant for the heavier states.  
This is very similar to the case of gravity, in which the higher-curvature invariants play no 
role in defining the properties of large black holes, whereas they give important corrections to Planckian black holes.

A purely Higgless Standard Model is incompatible with electroweak data, because it predicts a value of the custodial-symmetry breaking parameter $T$ of about $-0.3$, while present measurements indicate $T=0.07\pm 0.08$. However, the classicalized Higgsless Standard Model contains more dynamics at the scale $M_*$ than what described by the ordinary quarks, leptons, and gauge bosons. Whether we wish for it or not, the removal of the Higgs particle 
 from the Standard Model automatically introduces a tower of resonances,  with increasing mass (and spin) which, well above  $M_*$,  turn into classical states. These quantum resonances will give virtual contributions to electroweak observables. It is therefore natural to expect new effects on the parameters $S$ and $T$ of typical electroweak size. These effects cannot be computed without detailed knowledge of the dynamics at the scale $M_*$. All we can do here is to argue that it is possible for this dynamics to give a positive contribution to $T$ and an insignificant contribution to $S$, needed to reconcile the theory with experimental data.
  
  It is important to note the different roles played by the heavy and light resonances at various length scales.    
   The resonances that are heavier than $M_*$ do not affect the precision electroweak observables, since they represent classical states giving exponentially suppressed contributions to virtual processes, but  they are crucial in restoring unitarity at high energies. 
  On the other hand, resonances with mass around $M_*$  play 
  no role in scattering processes at energies much larger than $M_*$, but give important contributions
  to the electroweak precision parameters. In other words, since classical states
  decouple exponentially, the main contribution to electroweak observables comes from 
  quantum states in a narrow interval around $M_*$.    
   Although these quantum states must replace the contribution of a light Higgs in precision data,  this should occur just as a numerical coincidence, because their physical effect  is very different than the Higgs.  In the Standard Model, the Higgs is responsible for restoring unitarity at 
    arbitrarily high scales, whereas the quantum resonances in the Higgsless model play no role in 
    unitarizing processes at energies larger than $M_*$. The classical states, controlled by the infra-red sector of the theory, take care of this task.   
   
 The situation is somewhat analogous  to what happens in string theory with $g_s \sim 1$, with the classicalized states of the Higgless Standard Model playing the role of the string resonances.  Thus, from the phenomenological point of view, the classicalization may appear  as a UV-completion 
 by some sort of string theory, with the characteristic signature of a tower of string-type  resonances  above the scale $v$. 
 
 The occurrence of a tower of quantum states is reminiscent of technicolor, which also results in a series of  QCD string-type  resonances. However, there is a crucial difference. While in technicolor-like theories (like in any Wilsonian completion) the inclusive cross section decreases with $s$ well above the resonance region, in our case the total cross section keeps on growing with $s$ as larger (and more massive) classicalons are formed.

  \subsection{Using the Higgs as Classicalizer}
  
 Another interesting application of classicalization in the context of the Standard Model is the idea of maintaining the Higgs as a fundamental field, but using it as a classicalizing source.  The motivation for this approach is addressing the hierarchy problem by classicalization, whereas unitarity in WW-scattering is restored in the usual perturbative way through the Higgs field.   
    Since the Higgs is a scalar, it does produce  scalar gravity. So what we need is that the Higgs field classicalizes whenever integration is performed over  sources with momenta exceeding $M_*$. This can be achieved by adding operators ensuring that the Higgs is sourced by the energy-momentum of any Standard Model particle.  The simplest universal interaction is 
 \begin{equation}
 {\cal L}_{\rm int} \, =\,  { \kappa \over M_*^2}\,H^\dagger H\, T_{\mu}^{\mu}\, ,
   \label{higgscla}
   \end{equation}
  where $\kappa$ is a coupling constant and  $T_{\mu}^{\mu}$ represents the trace of the energy-momentum tensor of the Standard Model fields, evaluated order by order in $1/M_*$. 
 Although other types of interactions that couple the Higgs linearly can also be considered, we shall focus on the above interaction, which we consider to be the most straightforward choice.

 The classicalization effect of this coupling can be easily observed from the equation of motion of the Higgs field 
  in the presence of a localized source $T_{\mu}^{\mu} =  3\,  \theta(L-r)/L^4$, where $\theta$ is the Heaviside step function and the source is normalized  in such that its integrated value is $\int_0^{\infty} dr \, \, r^2 \, T_{\mu}^{\mu}=  1/L$. 
 Defining  $2H^\dagger H  =  [v+ \rho(r)]^2$, the equation for the Higgs in the presence of the source 
 takes the form
  \begin{equation}
  \left( \partial_r^2 +\frac{2}{r}\partial_r -\lambda^2v^2\right) \rho(r)-\frac{\lambda^2}{2} \left[ 3 v\rho^2(r) +\rho^3(r)\right] + ... = -\frac{3\kappa}{M_*^2L^4}\theta(L-r) \left[ v+\rho(r)\right] \, .
 \label{higggion}
 \end{equation}
 On the left-hand side we have left out the higher-order  self-couplings of the Higgs 
 that will come from the interaction of $H$ with its own energy momentum source.  Choosing the boundary conditions such that $\rho(\infty) = 0$, the asymptotic solution to eq.~(\ref{higggion}) for large $r$ and weak coupling is 
 \begin{equation}
 \rho(r) \, = \, \frac{\kappa v L_*^2} {r  L} \, {\rm e}^{-\lambda v r}  \, .
  \label{higgsionpm}
  \end{equation} 
  
    Depending on the sign of $\kappa$, we can distinguish two different regimes. 
 If $\kappa <0$, the source pushes  $\rho(r)$  towards negative values  
  as $r\rightarrow 0$,  but  $\rho$ cannot become smaller than  $-v$, because beyond that point the expectation value of the Higgs simply vanishes.  Thus, in such a case the classicalized configuration 
  (``Higgsion") represents a sphere of radius $r_*^{(-)}$, within which the Higgs expectation value vanishes, with 
   \begin{equation}
  r_*^{(-)} \, \simeq \, \frac{L_*^2}{ L} \, . 
  \label{minus}
  \end{equation}
  
  When $\kappa >0$,  $\rho(r)$ is driven towards positive values, and the expectation value 
    of the Higgs increases when we approach the source.  This growth is stabilized by higher-order self-interactions at the distance
  \begin{equation}
 r_*^{(+)} \, \simeq \, \frac{v L_*^3}{ L}  \, .
  \label{plus}
  \end{equation}  
    Note that  the growth could also be stabilized by the non-linearities in the Higgs potential but, for
  weak coupling ($\lambda  \ll  1$),  these are less important  than higher-order operators such as, for instance,   $L_*^2(\rho +v)^2 (\partial_{\mu} \rho)^2$. 

 The values of the $r_*$-radius in eqs.~(\ref{minus}) and (\ref{plus})  are legitimate as long as they are shorter than the Compton wavelength of the Higgs particle in the vacuum,   
  $m_H^{-1} \equiv (\lambda v)^{-1}$.  Beyond that value, $r_*$ does not further grow with energy. For the two cases, this translates into a condition on $\lambda$, 
  \begin{equation}
 \lambda \,  \ll  \,  {L \over L_*^2v}~ {\rm for}~\kappa <0 \, ~~~ {\rm and} ~~~\, 
 \lambda \,  \ll    \,  {L \over L_*^3v^2}~ {\rm for}~\kappa >0 \,. 
 \label{bounds}
 \end{equation}  
 In other words, classicalization requires a mild hierarchy between the Higgs mass 
 and the scale $M_*$.  Thus, if classicalization solves the hierarchy 
 problem at the TeV scale, a relatively light Higgs is preferred. 
 
    
        For  $\sqrt{s} \, \gg \, M_*$ the scattering of Standard Model particles is  universally dominated by  production of Higgsions, with total inclusive cross section 
  \begin{equation}
\sigma \sim   {r_*^{(-)}}^2  \simeq  sL_*^4   \, ~ {\rm for}~\kappa <0 \, ~~~ {\rm and} ~~~\, 
\sigma \sim   {r_*^{(+)}}^2  \simeq  sv^2L_*^6 \,
~ {\rm for}~\kappa >0 \,.  
  \label{sminus}
  \end{equation}
 This cross section grows with $\sqrt{s}$ until it freezes at the value  $\sigma \sim m_H^{-2}$. This happens at $\sqrt{s} \sim   1/(m_HL_*^2)$ for $\kappa <0$ and at $\sqrt{s} \sim   1/(m_HvL_*^3)$ for $\kappa <0$. 
  Above this scale the energy pumped into the system goes into creating 
heavier and heavier Higgsions,  without increasing their size (at least until the Higgsion becomes a true gravitational black hole at $\sqrt{s} \sim M_P$).

 As we have already mentioned,  classicalization requires the mass of the Higgs to be 
parametrically smaller than the classicalization scale $M_*$.  However, having the mass of the classicalizing field parametrically  one-loop factor  below the scale $M_*$ is enough for classicalization to occur, and a one-loop hierarchy between the Higgs mass and $M_*$ is perfectly natural.  The Higgs mass receives contributions from quantum loops, which are evaluated by integrating over sources of Standard Model particles (such as top--antitop pairs). Because of the interaction in eq.~(\ref{higgscla}), sources with energies $E>M_*$ will become classical and their decoupling factor will be at least 
   $\exp (-E^2/M_*^2)$. As a result, the total contribution to the Higgs mass will be given by 
$\delta m^2_{H}  \sim  {\rm (loop ~factor)} \times  M_*^2$. This is still consistent with having the Higgs as classicalizer at the scale $M_*$, since the Higgs mass is not preventing classicalization of sources with energies above $M_*$.

   We can now ask the following question about the couplings of the 
   Higgs to the energy-momenta of different Standard Model particles in the classicalizing interaction: how universal must they be? 
Flavor conservation restricts the form of such couplings, but  we wish to address the issue 
    from the point of view of classicalization.  At first sight, it seems that fermions that are weakly coupled to the Higgs (like first-generation quarks and leptons)  could be allowed to have weaker couplings also in the classicalizing interaction. For example, consider the Yukawa coupling of the Higgs to the electron,
    $g_eH\bar{e}e$. At one loop this coupling generates a quadratically sensitive Higgs mass  
    $\delta m_H^2 \sim (g_e^2/16\pi^2) \Lambda^2$.  In the presence of the classicalizing coupling 
    \begin{equation}
    {g^{\prime}_e}^{2}{H^\dagger H \over M_*^2}\bar{e}{\not\ \! \! \! \partial} e \, ,
    \end{equation}
  the electron loop will get classicalized and cutoff at the scale $\Lambda \sim M_*/g_e'$, which would not destabilize the weak scale as long as $g_e'  \simgt g_e$. Therefore, it seems that the electron can be allowed to have a much weaker classicalizing coupling than, say, a top quark.  But this consideration is  true only if we forbid
    all possible higher-dimensional operators that could couple the electron to the top quark generating unsuppressed contributions to the Higgs mass at some higher-loop level.  For instance, a four-fermion interaction  $\bar{e}e\bar{t}t/M_*^2$ would create a problem at two loops. Thus, if we allow for generic operators that couple different Standard Models
    species, then each of these species must have the same strength in the classicalizing interaction with the Higgs.   This implies that the production of Higgsions is a universal property of scattering processes with any 
 Standard Model particle.   Similarly, the decay of Higssions will be approximately universal, with roughly equal probability of producing any species of Standard Model particles.

 \section{Conclusions}
 
 Since the time of Rutherford's experiments, the exploration of smaller distances has been performed by means of particle probes of increasingly higher energies. Assuming that we could build accelerators with arbitrary high energies, we can ask the question of whether, as a matter of principle, this experimental strategy could be pursued {\it ad infinitum}. The answer is no. If the momentum transfer in a particle collision exceeds the Planck mass, the system collapses into a black hole and any information about shorter distances is inevitably cloaked inside its horizon. There is a minimum length beyond which no information can be extracted from collider experiments, no matter how powerful the accelerator is.
 
 In practice, this does not seem a serious limitation, since there is no immediate plan for a collider accelerating particles at Planckian energies. However, the situation could be very different if gravity becomes strong at the TeV scale~\cite{add}, since black holes could be formed in proton collisions at the LHC~\cite{bh}. In this paper we have argued that the inaccessibility of short distances at high energies -- a feature characteristic of gravity -- could also occur in certain non-gravitational non-renormalizable field theories and thus be relevant at much lower energies. Although we have not rigorously proved the phenomenon, which is intrinsically non-perturbative, we have given abundant evidence for its occurrence. 
 
 We have given several examples of theories which exhibit this phenomenon, called here {\it classicalization}. In these theories, the system of two particles colliding with energy and momentum transfer larger than $M_*$ (the mass scale associated with the non-renormalizable interaction) acts as a source for a field (called {\it classicalizer}), which consequently develops a classical configuration of size $r_*$, the {\it classicalon}. As the energy localized in the system increases, the radius $r_*$ grows. The classical size $r_*$ then becomes larger than any of the quantum lengths involved in the scattering process, and any information about the short-distance behavior of the theory is permanently enshrouded inside the extended classical object. 
 
Contrary to the case of black holes, the formation of classicalons does not imply the existence of horizons and therefore not all the short-distance information becomes inaccessible. Classicalons act only on special processes, whose nature depends on the structure of the underlying theory.

The phenomenon of classicalization suggests an alternative route in dealing with theories that seemingly violate unitarity in high-energy scattering processes. Instead of following the traditional Wilsonian approach and modifying the high-energy behavior of the theory by introducing new degrees of freedom, we can invoke classicalization as a cure. Some theories, which apparently violate unitarity, have in reality their own resources to bypass the problem at the non-perturbative level. As one approaches the energy scale of unitarity violation, large classical objects are formed and we do not need to specify the short-distance behavior of the theory to know its high-energy behavior. 

Classicalization could have dramatic consequences for theories at the electroweak scale. We have shown that the Standard Model, even without the Higgs, could have no clash with unitarity above the scale $v$, because of the classicalization of the longitudinal components of the gauge bosons. Alternatively, classicalization could solve the hierarchy problem of the Standard Model, if the Higgs field behaves as a classicalizer. In both cases, classicalization predicts sensational phenomena visible at the LHC.  
    
       We would like to stress that the essence of classicalization is at the heart of gravity as well as of string theory. In fact classicalization implements a physical mechanism in which the energy pumped into the system in order to localize it causes an increase of its effective size. This phenomenon naturally sets a fundamental length where the quantum/classical cross-over takes place\cite{GUP2}. Non-Wilsonian self-completeness by means of classicalization can only be achieved in theories possessing in their spectrum a classicalizer field\footnote{In string theory the classicalizer lies in the closed string sector and the existence of the classicalizer is guaranteed by the open/closed string correspondence.}. Similarly to what happens in gravity, the non-linear dynamics of the classicalizer field determines the effective size of any localized distribution of energy and sets the energy scale at which such a distribution turns into a classical configuration. 

  \vspace{5mm}
\centerline{\bf Acknowledgments}

We thank G. Gabadadze for useful discussions.
The work of G.D. was supported in part by Humboldt Foundation under Alexander von Humboldt Professorship,  by European Commission  under 
the ERC advanced grant 226371,  by  David and Lucile  Packard Foundation Fellowship for  Science and Engineering, and  by the NSF grant PHY-0758032. 
The work of C.G. was supported in part by the grants FPA 2009-07908, CPAN (CSD2007-00042), and HEPHACOS P-ESP00346.
The work of A.K is supported by a PEVE-NTUA-2009 grant. 
    
  \section*{Appendix: Stability Analysis} 
 Although, as previously pointed out, absolute stability is not really an issue, we would like to examine 
the stability of the classicaling configurations under linear perturbations. For this, let us 
consider the theory described by the action
\begin{eqnarray}
{\cal{L}}_\phi={1\over 2}(\partial_\mu\phi)^2-\frac{L_*^4}{4}  (\partial_\mu\phi)^4 \, .\label{f}
\end{eqnarray} 
 The  spherically symmetric  classicalizing  scalar configuration $\phi=\phi_0(r)$
 satisfies 
\begin{eqnarray}
\partial_r\phi_0+L_*^4(\partial_r\phi_0)^3=\frac{Q}{r^2}\, , ~~~  \nonumber 
\end{eqnarray}
where the charge $Q$ is localized within $r<L\ll L_*$. 
The solution to this equations has the following profile at short and large distances  ($r_*^2=QL_*^2$)
\begin{eqnarray} 
&&\phi_0\sim \frac{1}{r}\,  ~~~~r\gg r_*\, ,\nonumber \\
&&\phi_0\sim \frac{r^{1/3}}{r_*^{4/3}}\,  ~~~~r\ll r_* \, . \label{ff2}
\end{eqnarray}
 By setting $\phi=\phi_0+\varphi$, the dynamics of the perturbations $\varphi$ is described at  quadratic order by
\begin{eqnarray}
{\cal{L}}^{(2)}=\frac{1}{2}\left[1-L_*^4(\partial_\sigma \phi_0)^2\right]\partial_\mu \varphi\partial^\mu \varphi-L_*^4
\partial_\mu\phi_0\partial_\nu \phi_0\partial^\mu \varphi\partial^\nu \varphi \, .\label{fluct}
\end{eqnarray}
For the background (\ref{ff2}) and  for  $\varphi=\varphi_0(r) e^{-i \omega t}$, we find that  $\varphi(r)$ satisfies
\begin{eqnarray}
\frac{1}{r^2}\partial_r\left[r^2\Big{(}1+3 L_*^4 (\partial_r\phi_0)^2\Big{)}\partial_r\varphi_0\right]+
\omega^2\Big{(}1+ L_*^4 (\partial_r\phi_0)^2\Big{)}\varphi_0=0\, .
\end{eqnarray}
The solution for spherically symmetric fluctuations to the above equation describes 
 ingoing and outgoing plane waves at large  distances ($r\gg r_*$) from the 
classicalon,
\begin{eqnarray}
\varphi\sim \frac{e^{-i \omega (t\pm r)}}{r}\, , ~~~~r\gg r_*\, ,\nonumber 
\end{eqnarray}
whereas, close to the classicalon
\begin{eqnarray}
\varphi_0\sim C_1\,  r^{1/6}\, J_{1/6}\big{(}r\omega/\sqrt{3}\big{)}+C_2\,  r^{1/6}\, Y_{1/6}\big{(}r\omega/\sqrt{3}\big{)}
\, , ~~~~r\ll r_*\, .
\end{eqnarray}
Since $r^{1/6} Y_{1/6}, ~~r^{1/6} J_{1/6}$ remains finite for $r\to 0$, there are no growing modes to destabilize the
 classicalizer solution and the solution is stable (at least for linear perturbations). 

Similarly, classicalizing configurations induced by vectors are also stable. For example, we will show below that 
 the  theory described in  eq.~(\ref{caction}) :
\begin{eqnarray}
&&{\cal{L}}_A=-\frac{1}{4}\left(F_{\mu\nu}^2-\frac{1}{4} L_*^4 (F_{\mu\nu}^2)^2\right)\label{cl}
\end{eqnarray}
is stable under linear perturbations. 
Let us consider small fluctuations  $f_{\mu\nu}=\partial_\mu a_\nu-\partial_\nu a_\mu$ around a classical background 
 $F^{(0)}_{\mu\nu}$ such that
\begin{eqnarray}
F_{\mu\nu}=F^{(0)}_{\mu\nu}+f_{\mu\nu} \label{Ff}
\end{eqnarray}
Then, the quadratic action for $f_{\mu\nu}$ turns out to be
\begin{eqnarray}
{\cal{L}}^{(2)}=-\frac{1}{4}\left[ \left(1-{1\over 2}L_*^4{F^{(0)}}^2\right)\delta_{\kappa\mu}\delta_{\lambda\nu}-L_*^4
F^{(0)}_{\mu\nu}F^{(0)}_{\kappa\lambda}\right] \, \,  f^{\kappa\lambda}f^{\mu\nu}
\end{eqnarray}
On a static electric background 
$F^{(0)}_{0i}=E_i$, like the one described in sect.~\ref{sec3},  the action in terms of the perturbed electric ($e_i$) and magnetic ($b_i$) fields
\begin{eqnarray}
f_{0i}=e_i\, , ~~~f_{ij}=\epsilon_{ijk}b_k  \label{feb}
\end{eqnarray}
is written as
\begin{eqnarray}
{\cal{L}}^{(2)}={1\over 2}\left\{\left(1+L_*^4\, \vec{E}^2\right)\delta_{ij}+2 L_*^4 E_iE_j\right\}\,  e_ie_j- {1\over 2}
(1+ L_*^4
\, \vec{E}^2)\, {\vec{b}}^2\, .
\end{eqnarray}
 For the classicalizing background,  the above Lagrangian describes  ordinary electromagnetism, which however
 has  anisotropic 
electric susceptibility tensor for $r\ll r_*$, due to an effective polarization of the classicalon.

\end{document}